# R&D works for Superconducting Magnet for Future Accelerator Applications in Japan

**Toru Ogitsu[1a], Tatsushi Nakamoto[1], Ken-ichi Sasaki[1], Michinaka Sugano[1], Masami Iio[1], Kento Suzuki[1], Makoto Yoshida[2], Satoshi Awaji[3b], Naoyuki Amemiya[4c];**

[1] KEK Cryogenics Science Center, [2] KEK Institute of Particle and Nuclear Physics,
[3] Tohoku University High Field Laboratory for Superconducting Materials,
[4] Kyoto University Graduate School of Engineering Department of Electrical Engineering
[a] toru.ogitsu@kek.jp, [b] awaji@imr.tohoku.ac.jp, [c] amemiya.naoyuki.6a@kyoto-u.ac.jp

## Executive Summary

KEK cryogenics science center is developing superconducting magnet technology for future accelerator science. Three major technological categories are focused; 1) high precision 3D magnetic field technology based on the g-2/EDM magnet developments, 2) rad-hard superconducting magnet technology based on the COMET magnet developments, and 3) high magnetic field superconducting magnet technology for future colliders based on the LHC MQXA and HL-LHC D1 magnet developments. Extensive studies including $Nb_3Sn$ conductor and magnet developments for high field magnets and HTS for rad-hard superconducting magnets are ongoing with various collaboration programs such as the US-Japan research collaboration.

## 1. Introduction

R&Ds on superconducting magnets for accelerator applications have been conducted at KEK for more than 40 years. One of the major accomplishments is the development and the construction of LHC insertion quadrupole MQXA [1], which has the highest conductor field of 8.6 T. After the successful development of MQXA and their successful operation, KEK is now developing the large aperture beam separation dipole magnet D1 [2] for HL-LHC. KEK has also succeeded developments of superconducting accelerator magnets for its own projects such as superconducting beam line for T2K neutrino experiment facility [3] and developments of solenoid magnets for detector and muon experiment program such as COMET [4] and g-2/EDM at J-PARC [5].

Although these project-based magnets were developed with the NbTi superconductor, R&Ds of magnet technologies with advanced superconductor such as A15 or HTS have been conducted as well. For A15, $Nb_3Sn$ conductor development, aiming for the Future Circular Collider (FCC-hh) proposed in Europe, is now being conducted in collaboration with CERN, Tohoku Univ., Tokai Univ., NIMS, and two Japanese industrial partners [6]. For HTS developments, US-Japan collaboration is formed by KEK, Kyoto Univ., LBNL, and BNL for high field and high radiation environment applications [7]. The quench protection as well as effects of shielding current in the HTS conductors are also studied.

Based on the experiences next mid-term goal for the superconducting magnet technologies at KEK are defined in three categories; 1) high precision 3D magnetic field technology based on the g-2/EDM magnet developments, 2) rad-hard superconducting magnet technology based on the COMET magnet developments, and 3) high magnetic field superconducting magnet technology for future colliders based on the LHC MQXA and HL-LHC D1 magnet developments.

## 2. Status of KEK superconducting magnet development

### 2.1 High precision magnetic field

The J-PARC muon g-2/EDM experiment requires 3 T superconducting solenoid as muon storage magnet with high homogeneous magnetic field, less than +/- 0.1 ppm. To satisfy the requirement, new analytical code using truncated singular value decomposition method to optimize coil position and size is being newly developed [8], and design work is in progress using both the new code and existing commercial FEM code that can calculate with non-linear effect. Other than the high precision design technology, precise magnetic field control technology is also important, so advanced passive shimming technology using commercial MRI magnet is studied. These 3-D magnetic field design and control technologies are being developed in collaboration with Ibaraki University. Accompanying the precise magnetic field control, precise magnetic field monitoring system development is necessary. US-JP collaboration about NMR magnetometer with ultra-high precision [9] is being progressed effectively. High precision 3-D magnetic field control is also required for COMET transport curved solenoids. The solenoids require additional dipole field to separate particles by charge and select the beam energy. The solenoid assembly that accommodates with the small dipole corrector is developed for the current muon transport solenoid.

### 2.2 Rad-hard superconducting magnet

Development of rad-hard superconducting magnet technologies were started with the construction of the superconducting magnet beam line for J-PARC neutrino experiment, which was designed to transport the high intensity proton beam of about 1 MW. Series of gamma ray irradiation tests were performed on the organic materials used in the magnet system and rad-hard materials were selected according to each radiation levels.

Further studies were performed with the development of COMET superconducting solenoid system. The system includes the capture solenoid that contains the 50kW proton target in the aperture. The resultant radiation levels are order of higher than that of neutrino magnet system. The rad-hard organic material development was launched and various materials such as bismaleimido-triazin based GFRP were developed to achieve radiation hardness of about 100 MGy[10].

The degradation of RRR on high purity aluminum and copper due to neutron irradiation were also concerned. The degradation of RRR on stabilizer may affect to the magnet quench protection. For conduction cooling magnet, degradation RRR of pure aluminum thermal conductor reduces the thermal conductivity and increases the temperature of the coils. Series of irradiation studies were launched and the neutron irradiation to the aluminum and copper used for the superconductor stabilizer and pure aluminum used for the thermal conductor was performed. The irradiation was done at Kyoto University Research Reactor Institute (KURRI) with their neutron irradiation facility with cryogenics temperature. The test results indicates that both copper and aluminum show degradation of RRR with the irradiation level of about $10^{19}$ n/m$^2$. But in case of aluminum the degradations were 100 % recovered by thermal cycle to the room temperature while for copper recovery is only about 90 %. Considering the effect, the design and operation scenario of the COMET capture solenoid have been designed such that it can be operated stably for the lifetime of the experiment [11]. Since the degradation of the RRR may be influenced by the irradiation particles or their spectrum, several samples were attached to the capture solenoid that will enables to check the RRR degradation online.

**2.3 High field superconducting magnet technology**

One of the major accomplishments in the high field accelerator magnet at KEK is the development and the construction of LHC insertion quadrupole MQXA [1], which has the highest conductor field of 8.6 T. The technologies accumulated with the development of MQXA is now utilized for the development of HL-LHC D1 magnet. Although the peek field of the D1 magnet, 6.6 T, is much lower than that of MQXA, due to the large aperture and the dipole coil configuration, the accumulated electro-magnetic force results in the very high stress of about 100 MPa. This requires the high precision control of the coil pre-stress to manage the coil stress within the tolerable range during the assembly, cooldown, and excitation.

Those magnet development was based on the NbTi conductors, but there were also several developments with A15 conductors ($Nb_3Al$ or $Nb_3Sn$). For $Nb_3Al$ conductor with RHQT method, the conductor development was done in collaboration with NIMS and FNAL and the developed cable was used for sub-scale test coil and successfully tested under 10 T field. For $Nb_3Sn$ conductor the development is now under way in collaboration with CERN, Tohoku Univ., Tokai Univ., NIMS, and two Japanese manufacturers. One of the developed conductors has reached 1100 A/mm2 at 16 T, and the test production of 5 km level conductors with the similar specification is now in progress.

**3. Future projects and R&D plans at KEK**

**3.1 High precision magnetic field for accelerator sciences**

For high precision magnetic field magnet technology, the obvious near-term goal is to complete the development and construction of the g-2/EMD muon storage magnet. The development efforts are ongoing to achieve more efficient way to construct the magnet with optimum costs. Since the absolute precision of the magnetic field measurement is also critical to the success of the experiment, the NMR measurement system development is continued in collaboration with the US g-2 experiment team.

There is also a possibility to upgrade the experiment with the higher field storage magnet. Although the engineering studies are not yet started, future R&D may be anticipated in collaboration with the high field magnet development team.

Another project expected at J-PARC is COMET phase II project. In this project to improve the precision of the experiment, another curved solenoid system with dipole field is required. The system requires much larger aperture than the muon transport solenoid installed in phase I experiment and that more complex structure with curved dipole coil may be required. More sophisticated design that can construct the required 3D field will be required. Magnetic design method as well as mechanical structure engineering that can accommodate the complicated electro-magnetic force is required.

**3.2 Rad-hard superconducting magnet for future accelerator programs**

At J-PARC Material and Life Science Facility (MLF), a new proposal to build the second target station has arisen. For this target station muon production solenoid of about 1 T central field is directly attached to the 1 MW target that produce both muon and neutron. The facility aims to produce 50 to 100 times more muons than that of the current MLF muon source that results in world leading intense muon source. For the muon production solenoid, HTS conductor coil cooled by 20 K helium gas supplied from the refrigerator, of which is also supply refrigeration to the neutron moderator, is considered [12]. The solenoid requires high radiation hardness and also high reliability on quench protection.

To develop the rad-hard HTS magnet technologies, R&D programs are now on going with the US-Japan cooperation framework. With this R&D program, the series of irradiation tests are now on going and various ReBCO conductor such as GdBCO, EuBCO, or YBCO will be compared, and most sustainable conductor will be selected. There is also the development of new insulation material based on the sol-gel type of inorganic materials is also ongoing [13]. The quench protection studies are led by Kyoto University in collaboration with LBNL for HTS conductor with various stabilizers.

The interaction region magnets for future hadron collider such as FCC-hh will requires radiation hardness as well. The study on radiation hardness as well as mechanical durability in various operation cycles for the impregnation resin, in case of wind and react coil, should be performed. The other insulation material technologies such as sol-gel type of inorganic materials may be studied as well.

The currently on-going US-Japan cooperation obviously benefit to these projects and should be extended. The mid-term goal of this R&D is of course realization of the second target station with appropriate budget supply from Japanese government.

### 3.3 High field superconducting magnet technology for future colliders

One of the R&D targets can be the beam separation dipole (D1) for FCC-hh (Future Circular Collider hadron-hadron) that is 12 T 100 mm diameter aperture dipole magnet. Although the magnetic field is lower than the anticipated FCC arc dipole magnet field of 16 T, the large aperture results in higher force in the coil. Since mechanical stress management in the magnet structure as well as stress durability of the $Nb_3Sn$ conductor can be the important factor, study on magnet structure and mechanical properties of the conductor will be the major R&D target. For the mechanical properties of the conductor, measurement of the $I_c$ dependence with various mechanical stress can be performed. A development of mechanically strengthened $Nb_3Sn$ conductor can also be performed since such studies were already performed by some of the collaborators such as Tohoku Univ [13].

For further future, the above R&D can be extended to the R&D of 16~20 T accelerator magnet, by combining the 12 T $Nb_3Sn$ large aperture dipole with 4~8 T HTS (or $Nb_3Sn$) insertion coils. For the HTS part of the development not only high field applicability is required but also development of high current cable conductor should be performed. The current US-Japan cooperation already included the study on the high current cable development and the effort should be continued. In the collaboration with Kyoto Univ., extensive studies have been performed on influence of shielding current in HTS tape conductor [14]. The work will be essential to develop the magnet that can achieve the accelerator field quality. For that sense, the extension of the collaboration should be essential to the R&D. Since the current study on both $Nb_3Sn$ and HTS magnet technologies include the radiation hardness studies, the results of the R&D should lead to the high field, high radiation hard accelerator magnet technology required for FCC insertion quadrupole magnets.

### 4. Summary

KEK cryogenics science center is developing superconducting magnet technology for future accelerator science. Three major technological categories are focused; 1) high precision 3D magnetic field technology based on the g-2/EDM magnet developments, 2) rad-hard superconducting magnet technology based on the COMET magnet developments, and 3) high magnetic field superconducting magnet technology for future colliders based on the LHC MQXA and HL-LHC D1 magnet developments. Extensive studies including Nb$_3$Sn conductor and magnet developments for high field magnets and HTS for rad-hard superconducting magnet are ongoing with various research programs such as the US-Japan research collaboration.

It should be noted that another white paper is to be separately submitted, in cooperation with CERN, for superconducting detector magnet technology.